\documentstyle[12pt,psfig]{article}
\begin{document}

\title{
Simulating Plasma Turbulence in Tokamaks
}

\author{
Jeremy Kepner\\
Scott Parker\\
Viktor Decyk
}

\date{
\null
}

\maketitle

The development of magnetic fusion energy is a continuing national
priority. Present day toroidally confined plasma experiments (tokamaks)
are producing fusion power comparable to the input power needed to heat
the plasma ($P_{fusion} \approx 0.3 P_{in}$). One of the basic physics
problems is to understand turbulent transport phenomena, which cause
particles and energy to escape the confining magnetic fields. Plasma
turbulence is challenging from a theoretical point of view because of
the nonlinearity and high dimensionality of the governing equations.
Experimental measurements in the core region of a tokamak are limited by
the extremely high temperatures, ${\cal O} (10^8)$ Kelvin, of the
confined plasma. The high levels of both theoretical and experimental
difficulty highlight the potentially important role of numerical
simulations in developing a predictive model for turbulent transport.
Such a model would dramatically reduce uncertainties in tokamak design
and could lead to enhanced operating regimes, both of which would reduce
the size and expense of a fusion reactor.

An understanding of turbulent transport and the exploration of modes of
operation that suppress turbulence are central goals of the numerical
tokamak project, one of the Grand Challenges of the national HPCC
program. The process of modeling entails three main steps: (1) the
development of a simplified model of tokamaks that encompasses the
essential physics of the relevant instabilities, (2) the creation of
numerical algorithms for solving the governing equations, and (3)
implementation of these methods on massively parallel architectures.
This third step is necessary if we are to achieve simulations of
sufficient size and resolution to explain the trends seen in
experiments.

The state of the plasma is given by the distribution function
$f({\bf{x}},{\bf{v}},t)$, whose evolution is described by the
gyrokinetic equations---a reduced set of equations derived from the
Vlasov equation by phase averaging over the ion gyromotion while keeping
only the relevant temporal and spatial scales~\cite{ref:FriChe,ref:Lee}.
This averaging of the fast gyromotion reduces the dimensionality of the
governing equations from six to five. In addition, recently developed
numerical methods make it possible to follow only the perturbations of
the distribution function ${\delta f}$ from a stationary equilibrium
(see~\cite{ref:ParLee} and references therein). Even with the
considerable complexity of the gyrokinetic equations and the so-called
``$\delta f$ method,'' the algorithms are analogous to those used in
conventional particle-in-cell (PIC) simulations. PIC codes are both
memory- and CPU-intensive, and the effective use of high performance
computing, in particular massively parallel architectures, which require
a domain decomposition of the problem, is essential for the success of
the numerical tokamak project. Fortunately, this problem lends itself
naturally to a one-dimensional decomposition.

As an example of a few partially realized goals of the numerical tokamak
project, we present here some recent results for the simulation of modes
that may act as barriers to turbulent transport. Our three-dimensional
gyrokinetic simulation code is being used to study two effects that are
linearly stabilizing and that may cause the formation of transport
barriers: reversed magnetic shear and peaked density
profiles~\cite{ref:ParMynArtCumDecKepLeeTan}. We have found that weak or
negative magnetic shear, in combination with a peaked density profile
relative to the temperature profile, greatly suppresses turbulence in
the central region of the simulations. Similar features have been seen
experimentally~\cite{ref:LevEtAl}.

\subsection*{Tokamak Geometry}

\begin{figure}[tbh]
\centerline{\psfig{figure=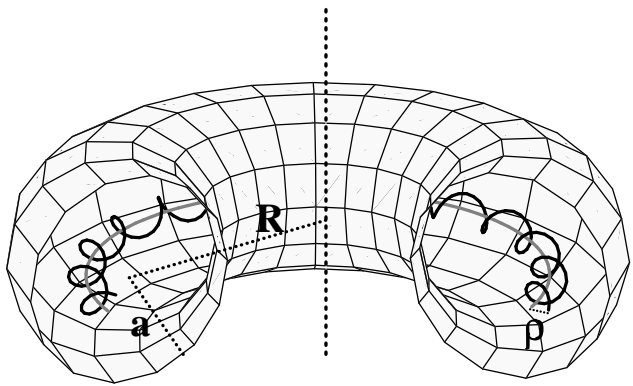,height=2.0in}}
\caption{Schematic drawing of a tokamak with major radius $R$ and minor
radius $a$ showing the path of a gyrating particle (black line) along
a magnetic field line (gray line) with a gyro radius $\rho$.}
\label{fig:Tokamak}
\end{figure}

The essential geometry of a tokamak is that of a torus defined by a
major radius $R$ and a minor radius $a$ (see Figure~\ref{fig:Tokamak}).
The ions within the plasma move rapidly around the torus, gyrating
tightly along the magnetic field lines, like rings on a wire. The radius
of the gyration is $\rho = v_t/\Omega$, where $v_t$ is the transverse
velocity and $\Omega = eB/mc$ is the gyro frequency, with $e$ being the
charge, $B$ the magnetic field strength, $m$ the particle mass, and $c$
the speed of light. The essential scale (resolution) of the system
(simulation) is set by $a/\rho$. Typically $R \approx 270$ cm, $a
\approx 85$ cm, and $\rho \approx 0.15$ cm.

\begin{figure}[tbh]
\centerline{\psfig{figure=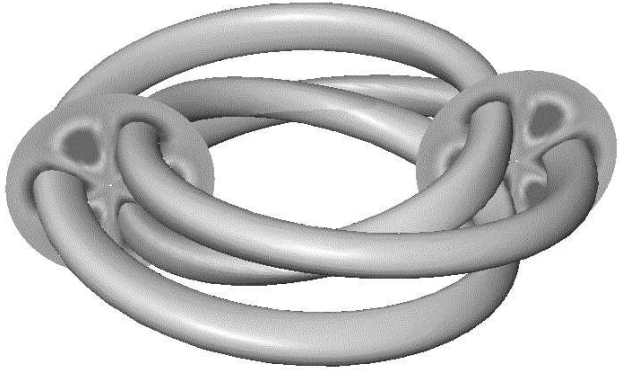,height=2.0in}}
\caption{Example of an $m=3$, $n=2$ mode in a torus with no shear
in the helical
magnetic field lines; $m$ is the poloidal mode number, and $n$ is the
toroidal mode number.
The isosurface shows the geometry of a typical
linear growth mode within the plasma; it also traces the magnetic field
lines.}
\label{fig:NoShear}
\end{figure}

The simplest arrangement of field lines is obtained by wrapping
current-carrying wires tightly around the minor axis of the torus,
creating straight magnetic field lines aligned with the torus.
Unfortunately, the magnetic field exerts a greater force on the inside
of the torus, which causes the ions in the plasma to drift across the
field lines. This problem can be partially alleviated by twisting the
field lines into a helical shape in such a way that the drifts
approximately cancel. A byproduct of the twisting is that the particle
trajectories become far more complex and are increasingly susceptible to
a wide range of instabilities, which tend to grow along the toroidal
modes of the tokamak. Figure~\ref{fig:NoShear} shows a simplified
example of a typical toroidal mode---an $m = 3$, $n = 2$ mode ($m$ is
the poloidal mode number, and $n$ is the toroidal mode number). This
could be, for example, the linear growth phase of an ion density
perturbation. The linear phase of an instability in realistic plasmas
typically has much larger values for $m$ and $n$ (10--100), but a
similar helical twist (the ratio $m/n$). In addition, the realistic case
has a sheared magnetic geometry (i.e., a radially varying amount of
helical twist in the magnetic field lines). Generally, the isosurfaces
of a perturbation follow the helical shape of the magnetic field lines.

\subsection*{Governing Equations}

The governing gyrokinetic equations are phase space conserving and have
the same form as the Vlasov equation:

\begin{eqnarray}
{\partial f \over \partial t} + {\dot{{\bf{z}}}} \cdot
{\partial f \over \partial {\bf z}} &=& 0 \nonumber
\end{eqnarray}

\noindent
where ${\bf{z}} = ({\bf R}, v_\parallel, \mu)$, with ${\bf R}$ being
the guiding center position of the gyrating particle, $v_\parallel$
the particle's velocity along the magnetic field line, and $\mu$ a
constant of motion that parameterizes the particle's velocity
perpendicular to the magnetic field.

The electrostatic toroidal gyrokinetic equations used for all the
results discussed here are those derived by Hahm~\cite{ref:Ham}. Because
the variations in $f$ are quite small, $f = f_0({\bf{z}}) + {\delta
f}({\bf{z}},t)$($\approx 1$\%) is used to carry out a perturbative
expansion. ${\delta f}$ in the expansion is solved by integrating the
characteristics of the resulting gyrokinetic equations. More details can
be found in~\cite{ref:ParLeeSan} and in references therein. It is
important to realize that although the solution of these equations is
algorithmically analogous to the solution of ${\bf F} = m{\bf a}$ plus
Maxwell's equations, much theoretical care and effort has been devoted
to simplifying the problem while retaining the important physics.
Simplifications include reduction of the dimensionality, elimination of
short space--time scales, and reduction of particle simulation noise.

\subsection*{Numerical Method and Parallel Implementation}

Particle-in-cell simulation has been used in the plasma physics
community for several decades. The general idea is that the particles
interact with both self-created and externally imposed electromagnetic
fields. The code thus has two distinct parts and data structures. For
illustration purposes, we consider the conventional electrostatic
equations, which retain the essence of the algorithm~\cite{ref:Dec_B}.
The trajectories of particles with mass $m$ and charge $q$ are given by

\begin{eqnarray}
d{\bf{v}}_i(t)/dt &=& (q/m) {\bf{E}}({\bf{x}}_i(t)) , \nonumber \\
d{\bf{x}}_i(t)/dt &=& {\bf{v}}_i(t) , \nonumber
\end{eqnarray}

\noindent
where the subscript $i$ refers to the $i$-th particle. The algorithm
generally used to solve this set of equations is a time-centered
leapfrog scheme,

\begin{eqnarray}
{\bf{v}}_i(t+dt/2) &=& {\bf{v}}_i(t - dt/2) + (q_i/m_i) {\bf{E}}({\bf{x}}_i(t)) , \nonumber \\
{\bf{x}}_i(t+dt) &=& {\bf{x}}_i(t) + {\bf{v}}_i(t+dt/2) , \nonumber
\end{eqnarray}

\noindent
where the electric field at the particle's position
${\bf{E}}({\bf{x}}_i)$ is found by interpolation from electric fields
previously calculated on a grid. The interpolation, a gather operation
that involves indirect addressing, accounts for a substantial part of
the computation time.

In the second part of the code, the fields created by the particles must
be found from Poisson's equation,

\begin{eqnarray}
\nabla^2 \phi &=& - 4 \pi \sum q n({\bf{x}}) \nonumber
\end{eqnarray}

\noindent
with ${\bf E}= - \nabla \phi$ and the sum being over each type of
particle species. Fourier transform methods are used to solve this
equation on a grid. Typically, the time for the field solver is not
large. The source term $n({\bf{x}})$ is calculated from the particle
position by an inverse interpolation,

\begin{eqnarray}
n({\bf{x}}) &=& \sum S({\bf{x}} - {\bf{x}}_i), \nonumber
\end{eqnarray}

\noindent
where $S$ is a particle shape function. This is a scatter operation that
also involves indirect addressing and consumes a substantial part of the
calculation time.

Since the dominant part of the calculation in a particle code involves
interpolation between particles and grids, it is important for a
parallel implementation that these two data structures reside on the
same processor. Different processors are assigned different regions of
space, and particles are assigned to processors according to their
locations~\cite{ref:LieDecdeB}. As particles move from one region to
another, they are moved to the processor associated with the new region.
Because particles must also access neighboring regions of space during
the interpolation, extra guard cells are kept in each processor, to be
combined or replicated as needed after the particles are processed. The
passing of particles from one processor to another is performed by a
particle-manager subroutine. The passing of field data between guard
cells is performed by a field-manager subroutine.

The field solver uses Fourier transform methods. There are two FFTs per
timestep, one for the charge density and one for the electric potential.
A parallel complex-to-complex FFT was developed with a transform method
in which the coordinate local to the processor is transformed first, the
data are then transposed so that the coordinate that was distributed
becomes local, and, finally, the remaining coordinate is transposed. The
maximum number of processors that can be used is limited by the maximum
number of grid points in any one coordinate, but this is not a severe
constraint at present since the numerical tokamak is designed for
systems in which this number is about 512.

During the transpose phase of the FFT, each processor sends one piece of
data to every other processor. This can be accomplished in a number of
ways, but the safest is always to have one message sent, one message
received, and so on. Flooding the computer with large numbers of
simultaneous messages tends to overflow system resources and is not
always reliable.

The structure of the main loop of the simplified code is summarized as
follows:

\newcounter{ijk}
\begin{list}{\arabic{ijk}. }{
\usecounter{ijk}
\setlength{\itemsep}{0.0in}
\setlength{\parsep}{0.0in}
\setlength{\topsep}{0.0in}
\setlength{\partopsep}{0.0in}}
\item
Particle coordinates are updated by an acceleration subroutine.
\item
Particles are moved to neighboring processors by a particle-manager
subroutine.
\item
Particle charge is deposited on the grid by a deposit subroutine.
\item
Guard cells are combined by a field-manager subroutine.
\item
Complex-to-complex FFT of charge density is performed by an FFT
subroutine.
\item
Electric potential is found in Fourier space by a Poisson solver
subroutine.
\item
Complex-to-complex FFT of electric potential is performed by an FFT
subroutine.
\item
Electric potential is distributed to guard cells by a field-manager
subroutine.
\end{list}
\medskip

This structure has the beneficial feature that the physics modules,
items 1, 3, and 6, contain no communication calls (except for a single
call to sum energies across processors). These modules can easily be
modified by researchers who do not have special knowledge of parallel
computing or message passing. The communications modules, items 2, 4,
and 8, handle data management but do not perform any calculation, and
can be used by physicists as black boxes, where only the input and
output must be understood. The FFT, items 5 and 7, are the usual
sequential FFTs, with an additional embedded transpose subroutine, which
can also be used as a black box. Furthermore, since most message-passing
libraries are quite similar, moving from one distributed-memory computer
to another simply involves replacing the message-passing calls in the
communications modules with new ones.

\subsection*{General Behavior}

\begin{figure}[tbh]
\centerline{\psfig{figure=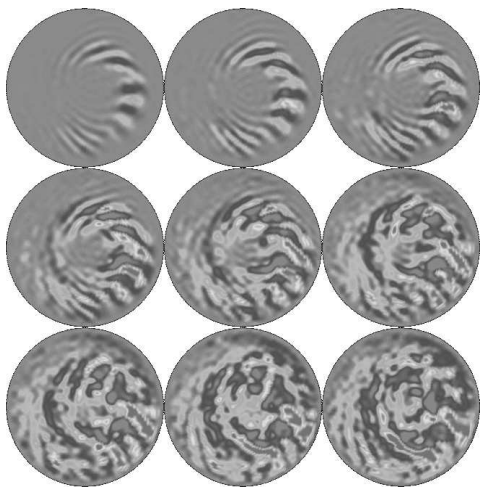,height=2.5in}}
\caption{Time evolution of ion density in a simulation
of tokamak plasma showing transition to a turbulent state.
Time sequences advance from left to right, top row to bottom row.}
\label{fig:TimeEvolution}
\end{figure}

When a simulation starts with a very small initial perturbation, the run
generally has two identifiable phases. The first is the linear phase,
where modes (i.e., standing waves) grow exponentially.
Lower-dimensional, time-independent eigenvalue techniques that are
fairly well understood theoretically can also be used to find linear
modes. The second phase is the turbulent stationary state, during which
the growth of the dominant linear modes saturates and the system settles
down to a statistical steady state. The transition from linear to
turbulent behavior is demonstrated in Figure~\ref{fig:TimeEvolution}.

\subsection*{Parallel Performance}

In our implementation, we use a one-dimensional domain decomposition
along the toroidal axis, which is generally more efficient and
significantly easier to program than a full three-dimensional
decomposition. In addition, the number of particles per processor
remains relatively constant along this axis, which minimizes load
imbalance. Furthermore, since the mean flow of the particles has been
subtracted off in the perturbation expansion, the particles themselves
do not move rapidly across cells, which keeps the communication low. For
these reasons, we would expect to get excellent parallel performance.

In the most recent efforts, Fortran 77 and the PVM message-passing
library have been used to port the code to the Cray
T3D~\cite{ref:Dec_A}. Production runs on the T3D show a performance of
approximately 14.4 MFlops per processor (10\% of the theoretical peak of
150 MFlops, which is typical for applications of this type). To test the
scalability of the code, 12 runs were timed with different numbers of
processors and problem sizes. These times, along with the parameters of
the runs, are shown in Table~\ref{tab:Timings}.

\begin{table}[tbh]
\begin{center}
\begin{tabular}{cccccc}
\hline
No. of & grid & particles & No. of  & wallclock & speed (msec-processor \\
particles & size & /grid cell & processors & sec/step & /particles-step) \\
\hline
$2^{13}$ &  $16^3$ & $2$ &  $16$ & $0.12$ & $0.23$ \\
$2^{14}$ &  $16^3$ & $4$ &  $16$ & $0.19$ & $0.19$ \\
$2^{15}$ &  $16^3$ & $8$ &  $16$ & $0.34$ & $0.16$ \\
$2^{16}$ &  $32^3$ & $2$ &  $32$ & $0.44$ & $0.21$ \\
$2^{17}$ &  $32^3$ & $4$ &  $32$ & $0.78$ & $0.19$ \\
$2^{18}$ &  $32^3$ & $8$ &  $32$ & $1.41$ & $0.17$ \\
$2^{19}$ &  $64^3$ & $2$ &  $64$ & $1.91$ & $0.23$ \\
$2^{20}$ &  $64^3$ & $4$ &  $64$ & $3.37$ & $0.20$ \\
$2^{21}$ &  $64^3$ & $8$ &  $64$ & $6.35$ & $0.19$ \\
$2^{22}$ & $128^3$ & $2$ & $128$ & $9.16$ & $0.27$ \\
$2^{23}$ & $128^3$ & $4$ & $128$ & $17.0$ & $0.25$ \\
$2^{24}$ & $128^3$ & $8$ & $128$ & $32.9$ & $0.25$ \\
\hline
\end{tabular}
\end{center}
\caption{T3D timings for different numbers of
processors and problem sizes.}
\label{tab:Timings}
\end{table}

\begin{figure}[tbh]
\centerline{\psfig{figure=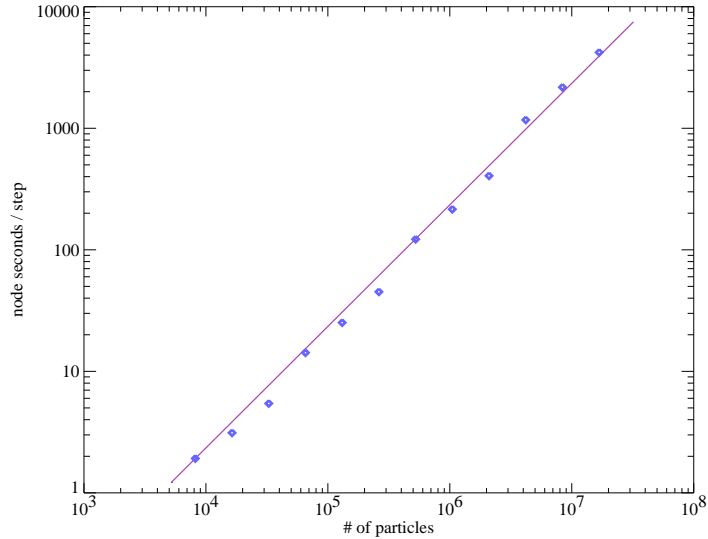,height=3.0in}}
\caption{Scaling of the code with problem size and number of processors.
The straight line indicates the expected time from a simple scaling
up of the smallest simulation.}
\label{fig:Scaling}
\end{figure}

To first order, the problem size is given by the total number of
particles. The total amount of computer resources consumed is the time
per step multiplied by the number of processors. These two quantities
are plotted in Figure~\ref{fig:Scaling}. In a perfectly scalable code,
which is represented by the straight line in Figure~\ref{fig:Scaling},
the resource consumption should be proportional to the problem size.
What is most impressive is that the full code, with all diagnostics and
outputs, was used to obtain these results. Simply put, this means that
doubling the number of CPUs or halving the size of the problem will
halve the computation time. In addition, it is worth mentioning that the
code has a very low communication/computation ratio ($< 0.01$). This
ratio is a rough measure of the fraction of time spent waiting for
processors to transmit information, and the low value for our program is
an indication that it can be expected to perform well if even more
processors or faster processors are used. This is encouraging because
the Cray T3E will have processors that are significantly faster.

\subsection*{Simulating Transport Barriers}

\begin{figure}[tbh]
\centerline{\psfig{figure=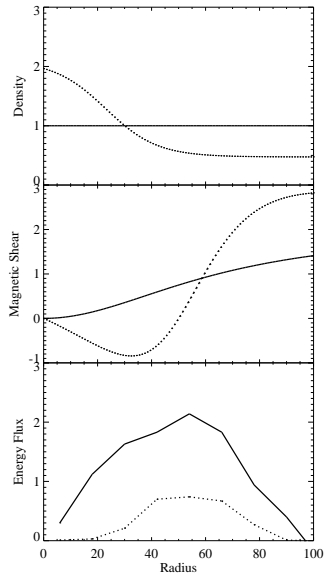,height=3.5in}}
\caption{Density (top), magnetic shear (middle), and energy
flux (bottom) profiles for the case with (solid line)
and without (dotted line)
peaked density and reversed magnetic shear profiles.
A large
decrease in the energy flux can be seen in the reversed shear case.}
\label{fig:Profiles}
\end{figure}

Transport barriers in tokamaks have been an important aspect in a host
of operational modes, such as the H-mode, where edge transport is
thought to be greatly reduced through poloidal shear flow. In the recent
enhanced reverse shear experiments on TFTR~\cite{ref:LevEtAl}, it has
been reported that density and ion heat transport are below conventional
neoclassical levels in the core. Comprehensive linear
calculations~\cite{ref:KesManRewTan,ref:RewTan} show this region to be
locally stable to micro-instabilities. New gyrokinetic simulations
presented here show that the combined effects of reversed magnetic shear
and a peaked density profile allow for a good confinement zone in the
core region~\cite{ref:ParMynArtCumDecKepLeeTan}.
Figure~\ref{fig:Profiles} shows the difference in energy flux for
simulations with and without reversed magnetic shear and a peaked
density profile. These new results are differ with past simulations
(that did not include these effects), which have generally shown a
global (slow) relaxation of the temperature profile.

\subsection*{Future Work}

Future directions for this work include the incorporation of more
detailed physics, such as an electron model that includes a trapped
component and studies of the effects of magnetic perturbations.
Ultimately, the goal is to understand plasma turbulence at a level that
is detailed enough to allow quantitative predictions of heat transport.
This will reduce uncertainties in design, and hence the cost, of future
tokamak reactors. As shown here, progress toward this long-term goal can
be made through close interaction between theory and direct numerical
simulation.

\subsection*{Acknowledgments}

The authors especially thank J.~Cummings, W.W.~Lee, H.~Mynick,
R.~Samtaney, E.~Valeo, and N.~Zabusky.
Much of this work was carried out at the
Princeton Plasma Physics Laboratory as an active part of the
community-wide Numerical Tokamak Project supported through the HPCC
Initiative.
Computing resources were provided by JPL at Caltech,
ACL at LANL, NERSC at LLNL, and the Pittsburgh Supercomputing
Center.
This work is supported by the U.S. Department of Energy, through
Contract DE-AC02-76CHO-3073, Grant DE-FG02-93ER25179.A000, and the
Computational Science Graduate Fellowship Program.

\noindent
Jeremy Kepner ({\tt jvkepner@astro.princeton.edu})
is a graduate student in the Department of Astrophysics at
Princeton University.
Scott Parker ({\tt sparker@buteo.colorado.edu})
is a professor in the Department of Physics at the
University of Colorado in Boulder.
Viktor Decyk ({\tt vdecyk@pepper.physics.ucla.edu})
is a research scientist in the Department of Physics at the
University of California in Los Angeles.


\begin{thebibliography}{99}

\bibitem{ref:Dec_A}
{\sc V.K. Decyk},
{\em How to write (nearly) portable Fortran programs for
parallel computers},
Computers in Physics,
7(1993),
pp. 418--424.

\bibitem{ref:Dec_B}
{\sc V.K. Decyk},
{\em Skeleton PIC codes for parallel computers},
Computer Physics Communications,
87(1995),
pp. 87--94.

\bibitem{ref:FriChe}
{\sc E.A. Frieman and L. Chen},
{\em Nonlinear gyrokinetic equations for low-frequency electromagnetic
waves in general plasma equilibrium},
Physics of Fluids,
25(1982),
pp. 502--507.

\bibitem{ref:Ham}
{\sc T.S. Hahm},
{\em Nonlinear gyrokinetic equations for tokamak microturbulence},
Physics of Fluids,
31(1988),
pp. 2670--2673.

\bibitem{ref:KesManRewTan}
{\sc C. Kessel, J. Manickam, G. Rewoldt, and W.M. Tang},
{\em Improved plasma performance in tokamaks with negative magnetic shear},
Physical Review Letters,
72(1994),
pp. 1212--1215.

\bibitem{ref:Lee}
{\sc W.W. Lee},
{\em Gyrokinetic particle simulation model},
Journal of Computational Physics,
72(1987),
pp. 243--269.

\bibitem{ref:LevEtAl}
{\sc F.M. Levinton, et.al.},
{\em Improved confinement with reversed magnetic shear in TFTR},
Physical Review Letters,
75(1995),
pp. 4417--4420.

\bibitem{ref:LieDecdeB}
{\sc P.C. Liewer, V.K. Decyk, and A. de Boer},
{\em A general concurrent algorithm for plasma particle-in-cell
simulation codes},
Journal of Computational Physics,
85(1989),
pp. 302--322.

\bibitem{ref:ParLeeSan}
{\sc S.E. Parker, W.W. Lee and R.A. Santoro},
{\em Gyrokinetic simulation of ion temperature gradient driven
turbulence in 3D toroidal geometry},
Physical Review Letters,
71(1993),
pp. 2042--2045.

\bibitem{ref:ParLee}
{\sc S.E. Parker and W.W. Lee},
{\em A fully nonlinear characteristic method for gyrokinetic simulation},
Physics of Fluids B,
5(1993),
pp. 77--86.

\bibitem{ref:ParMynArtCumDecKepLeeTan}
{\sc S.E. Parker, H.E. Mynick, M. Artun, J.C. Cummings,
V. Decyk, J.V. Kepner, W.W. Lee, and W.M. Tang},
{\em Radially global gyrokinetic simulation studies of transport barriers},
Physics of Plasmas,
3(1996),
pp. 1959--1966.

\bibitem{ref:RewTan}
{\sc G.W. Rewoldt and W.M. Tang},
private communication,
1995.

\end{thebibliography}
\end{document}